\documentclass[11pt,preprint,graphicx]{aastex63}

\usepackage{newtxtext,newtxmath}
\usepackage[T1]{fontenc}
\usepackage{ae,aecompl}
\usepackage{threeparttable}

\usepackage{amsfonts} % \mathbb \forall
\usepackage{amstext}
\usepackage{amssymb} % ulteriori simboli (ex: \blacksquare)
\usepackage{graphicx}
\usepackage{lineno}
\newcommand{\hi } {{\rm H}\,{\small\rm I} \,}
\def\etal{\it et al. \rm }

\begin{document}

\title{The Baryonic Tully-Fisher Relation I: WISE/Spitzer Photometry}

\author{Francis Duey}
\affiliation{Department of Astronomy, Case Western Reserve University, Cleveland, OH 44106}
\author[0000-0003-2022-1911]{James Schombert}
\affiliation{Department of Physics, University of Oregon, Eugene, OR 97403}
\author[0000-0002-9762-0980]{Stacy McGaugh}
\affiliation{Department of Astronomy, Case Western Reserve University, Cleveland, OH 44106}
\author[0000-0002-9024-9883]{Federico Lelli}
\affiliation{Arcetri Astrophysical Observatory (INAF), Florence, Tuscany, IT}

\begin{abstract}

\noindent We present WISE W1 photometry of the SPARC ({\it Spitzer} Photometry and
Accurate Rotation Curves) sample.  The baseline of near-IR fluxes is established for
use by stellar mass models, a key component to the baryonic Tully-Fisher relation and
other kinematic galaxies scaling relations.  We focus this paper on
determination of the characteristics of the W1 fluxes compared to IRAC 3.6$\mu$m
fluxes, internal accuracy limitations from photometric techniques, external accuracy
by comparison to other work in the literature and the range of W1 to IRAC 3.6$\mu$m
colors.  We outline the behavior of SDSS $g$, W1 and IRAC 3.6 colors with respect to
underlying SED features.  We also note a previously unknown correlation between WISE
colors and the central surface brightness, probably related to the low metallicity of
low surface brightness dwarfs.

\end{abstract}


\section{Introduction}

The classic Tully-Fisher (TF) relation links the rotation velocity of a disk galaxy
to its stellar mass and/or luminosity in a given photometric band (see Kourkchi \etal
2020). Because the observed rotation velocities do not depend on galaxy distance $D$,
while stellar luminosities depend on $D^2$, the TF relation has, historically, played
a crucial role in constraining the value of $H_o$ (Tully \& Fisher 1977, Sakai
\etal 2000).  The classic TF relation, however, breaks down at stellar masses below
approximately 10$^9$ M$_\odot$, when dwarf galaxies in groups and in the field
environment become progressively more gas rich.  By replacing the stellar mass with
the total baryonic mass (stars plus gas, $M_b$), one recovers a single linear
relation: the so-called baryonic Tully-Fisher relation (bTFR, Freeman 1999; McGaugh
\etal 2000; Verheijen 2001, Zaritsky \etal 2014).

The bTFR samples deeper into the galaxy mass function, as low-mass dwarfs typically
have high gas fractions and the neutral gas can constitute 80 to 90\% of the total
baryonic mass (Bradford \etal 2015).  This results in a large amount of scatter in
the classic TF relation with a corresponding loss in accuracy as a distance
indicator, which can be minimized by including the gas mass. Currently, the bTFR
extends over two decades in velocity and six decades in $M_b$ (McGaugh 2012, Iorio
\etal 2017).  Moreover, it displays a surprisingly small scatter, considering the
number of possible competing astrophysical processes that produce this relation
(Lelli, McGaugh \& Schombert 2016).  However, on the high baryonic mass end of the
bTFR, the stellar mass of a disk galaxy dominates.  Thus, while the range in the bTFR
will not be dramatically increased on the high mass end compared to the low mass end,
the linearity of the bTFR is highly dependent on accurate stellar masses at the high
mass end (Duey, Tosi \& Schombert 2023).

The current highest quality data to study the baryonic component of rotating galaxies
is the SPARC ({\it Spitzer} Photometry and Accurate Rotation Curves) dataset (Lelli
\etal 2016).  SPARC uses deep IR imaging to 1) determine the total stellar mass
through IR photometry and 2) compute the stellar gravitation contribution to the
observed rotation curve 
as a function of galaxy radius.  The first aspect is used to derive
the baryonic Tully-Fisher relation, the second aspect results in the new radial
acceleration relation (i.e. the RAR, Lelli \etal 2017).

For stellar mass determination, the near-IR offers a portion of a galaxy's spectrum
that is dominated by starlight free from strong emission lines and star formation
effects.  The low Galactic and internal galactic extinction make near-IR fluxes
more consistent across morphological types.  In addition, an optical to near-IR color
fixes a unique value for the mass-to-light ratio ($\Upsilon_*$) based on stellar
population models (Schombert, McGaugh \& Lelli 2022) converting near-IR fluxes into
stellar mass.  

The SPARC project depended on pointed observations from the {\it Spitzer} Space
Telescope (Werner \etal 2004). But, with the termination of the {\it Spitzer}
mission, future observations will be dependent on the all-sky dataset from the WISE
(Wide-field Infrared Survey Explorer) mission (Wright \etal 2010).  The goal of this
paper is to link the photometry system of {\it Spitzer} to WISE in order to determine
total IR luminosity and explore the meaning of galaxy colors with respect to the WISE
filter W1.  Both characteristics are key inputs into the stellar population models
that are used to convert luminosity in the baryonic component that is stellar mass.
In addition, the competing {\it Spitzer} and WISE photometry datasets provide an
opportunity to examine the astrophysics underlying galaxy luminosity at 3.5$\mu$m and
the uncertainties in galaxy photometry at these wavelengths for future studies of the
distance scale of galaxies (Schombert \etal 2020).

\section{Photometry}

\subsection{The SPARC Galaxy Database}

The SPARC dataset consists of high quality \hi rotation curves accumulated over
the last three decades of radio interferometry combined with deep near-IR photometry
from the {\it Spitzer} 3.6$\mu$m IRAC camera (Lelli, McGaugh \& Schombert 2016).
This provides the community an important combination of extended \hi rotation curves
(mapping the galaxy gravitational potential out to large radii) plus near-IR surface
photometry to map the stellar component (see also Zaritsky \etal 2014).  In addition,
the \hi observations also provide the \hi gas mass that, when corrected for small
amounts of He and heavier elements, becomes the total gas mass of a galaxy.  The
combined stellar and gas components is the total baryonic mass of a galaxy.

The SPARC sample spans a broad range in baryonic mass (10$^8$ to 10$^{11}$
$M_{\odot}$), surface brightness (3 to 1000 $L_{\odot}$ pc$^{-2}$) and rotation
velocity ($V_{f}$ from 20 to 300 km sec$^{-1}$).  The SPARC dataset also contains
every Hubble morphological type from S0 to Irr producing a representative sample of
different types of galaxies from dwarf irregulars to massive spirals with large
bulges.  Details of this sample are listed in Schombert \& McGaugh (2014) and the key
science results outlined in McGaugh, Lelli \& Schombert (2016).  The SPARC sample,
and analysis with respect to the bTFR, are presented in Lelli, McGaugh \& Schombert
(2016).  The analysis presented herein follows that paper with respect to error
analysis plus small additions and corrections to the data as outlined in Lelli \etal
(2019).  Our larger goal is to expand our near-IR photometry datasets in
preparation of the next generation of the SPARC sample from ongoing HI rotation
curves studies.  As discussed below, in order to compare WISE versus {\it Spitzer}
photometry, we have isolated a subset of 111 galaxies from the SPARC sample that
follow the quality criterion outlined in Lelli \etal (2019) plus have matching WISE
and IRAC images free of contaminating bright stars or neighboring galaxies.

\subsection{SPARC Photometry Pipeline}

With respect to exploring the stellar mass properties of galaxy with kinematic
information, such as a rotation curve, there are two critical aspects to the data
reduction process.  The first is the assignment of a total stellar mass of the galaxy
based on a total luminosity at some wavelength that minimizes the uncertainty in
going from luminosity to stellar mass.  For stellar population reasons, this is best
obtained in the near and mid-IR filters (but not so long in wavelength to encounter
star-forming hot dust features in the far-IR, Brown \etal 2014).  This is primarily
the domain of space imaging with its low IR background.  In comparison, UV and
optical wavelengths suffer from the short-lived effects of star formation and produce
highly color dependent, and increasingly uncertain, estimates of the mass-to-light
ratios (Taylor \etal 2011).

A second data reduction goal is the determination of stellar mass surface density as
a function of radius for a point-by-point comparison between baryon mass and the
observed rotation curve at that radius (the so-called radial acceleration relation,
Lelli \etal 2017).  This is the technique of surface photometry (Okamura 1988), the
determination of luminosity density along some isophotal shape.  The shape of choice
is an ellipse, motivated by the empirical observation that
ellipses are remarkably good describers of even the most irregular shaped galaxy
(Schombert \etal 1992).

The two forms of galaxy photometry, total and isophotal, are coupled as one uses the
ellipses defined by a galaxy's surface photometry to determine an isophotal magnitude
and an extrapolation of those same ellipses can result in a curve of growth to
determine a total luminosity (Schombert 2011).  In addition, other scale parameters,
such as disk/bulge scale lengths and half-light radii, are also extracted from the
surface photometry to quantify galaxy structure and mean density parameters such as
central surface brightness.

While the field of galaxy photometry has a long history (Peletier 2013), there are
two primary complications to obtaining accurate surface photometry (and isophotal
magnitudes) worthy of particular attention. The first is the problem of low
signal-to-noise with respect to the background.  Even the brightest galaxies fade to
background level at their edges.  Depending on a galaxy's light profile, galaxies can
have a significant amount of their flux in large, low S/N apertures. The
determination of an accurate total luminosity will require capturing or estimating
that flux.  The second complication is the contamination of a galaxy's luminosity by
foreground stars and nearby galaxy envelopes.  This is a particular problem in the IR
as the number of point sources are factors of ten higher than in the optical
(Schombert \& McGaugh 2014, Jarrett \etal 2019).  Various photometry programs have
dealt with this issue using different techniques ranging from doing nothing to masking
point sources to removing point sources with a PSF algorithm (see early work by
Jarrett \etal 2000 and a comprehensive attack on this problem by the GAMA project,
Wright \etal 2018).

Various galaxy photometry packages exist for the community (see Munoz-Mateos \etal
2015 for a detailed description of {\it Spitzer} galaxy photometry 
and Trujillo \etal 2020 for precision analysis of galaxy sizes).  Their
outputs differ depending on science goals but typically all have a common objective
of parametrizing either the total luminosity or some fraction of the total luminosity
that can be scaled by morphological type (Sandage \& Perelmuter 1990).  The
photometry for this project used an expanded version of the ARCHANGEL galaxy analysis
package (Schombert 2011, Schombert \& McGaugh 2014).  The ARCHANGEL package was
originally designed to analyze optical images of low surface brightness galaxies
(Schombert \etal 1992), but has many features that are well suited to near-IR
imaging, such as special algorithms adjusted for irregular galaxy morphology and low
S/N with respect to the sky.  Key to the ARCHANGEL package is a fast least-squares
ellipse fitting procedure that simultaneously fits and cleans isophotal regions.
Cleaning is accomplished by two procedures; 1) a manual masking of deeply embedded
stellar sources within the galaxy itself and 2) automatic masking of stars and
artifacts in the outer regions by a threshold algorithm.  Masked regions are replaced
by averaged values from the fitted ellipses for aperture photometry, but left masked
for surface photometry evaluation (which uses a mean isophotal value rather than a
total flux).

While masking is optimal for smooth, regular galaxies, such as ellipticals, it
becomes problematic for highly irregular late-type system with strong star formation
which produce clumps and knots that are hard to distinguish from foreground stars
(see Fig. 2, Schombert \& McGaugh 2014).  Surface brightness profiles are extracted
from the mean intensities around each fitted ellipse minus the masked regions.
Inhibitors to the pipeline prevent sharp changes in ellipse position angle, isophote
center and eccentricity which enhances an accurate transition from bulge to disk in
early-type disks.  If a set error threshold is exceeded, the previous fitted ellipse
is used as a default.  At the 1\% sky threshold fitting is halted, but ellipses are
continued to be evaluated to the edge of the frame.  These outer ellipses can be
compared to sky values obtained by histogram analysis or sky boxes and provide a
direct comparison of the uncertainty in the true sky value.

On the assumption that all of the rotating galaxies in our sample are oblate disks,
the semi-major axis is used as the profile radius.  Error bars are assigned based on
two uncertainties, the standard deviation in pixel intensities around the ellipse and
the error in the sky value.  For inner isophotes, the standard deviation dominates
the error budget.  For outer isophotes, the knowledge of the correct background
intensity dominates the error, which includes a combination of the flatness of the
images as well as the standard deviation on the mean of the sky boxes.  For space
imaging, frame flatness is rarely a problem and these sky values are also compared
with gaussian fits to the border pixel values (i.e., a histogram sky value).  Error
bars are assigned through a quadrature average of all the possible errors.

Most photometry projects assign a total magnitude using the Kron system, or some
isophotal equivalent (Kron 1980).  This technique determines a luminosity weighted
radius to define an elliptical aperture where the total flux is summed (see for
example an analysis of the SDSS Petrosian magnitudes, Schombert 2016).  Often a
masking routine has removed non-galaxy pixels from inside the aperture; some
techniques replace them with nearby intensity values, others just leave them to be
ignored.  Depending on the steepness of the galaxy's luminosity profile, a Kron
magnitude will capture between 85 to 95\% of the total luminosity (Schombert 2016,
Bertin \& Arnouts 1996) which is similar to the Holmberg magnitude outlined in
Schombert (2018).  A more detailed treatment of how a galaxy's luminosity profile
relates to its total luminosity can be found in Graham \& Driver (2005).

The photometric technique for this study also evaluates isophotal magnitudes, but
primarily uses curves of growth to determine total luminosities (Schombert \& McGaugh
2014).  In addition, metric apertures are used for comparison of colors (apertures of
a set arcsec size) and surface brightness profiles are used for radial mass profiles.
Metric magnitudes are generated by selecting an isophotal radius and summing the
pixel fluxes inside that elliptical aperture. However, an isophotal magnitude has
variable meaning and does not typically result in an accurate total flux even with
aperture corrections (see below).  Curves of growth are constructed using the cleaned
images and the surface photometry fits.  Our procedure differs slightly from raw
curves of growth (where one simply sums the flux in each aperture) by using a partial
pixel routine on the repaired images (masked pixels replaced with ellipse
intensities) and the outer apertures have their fluxes calculated using both the raw
pixel values and a mean annular flux defined by the surface brightness at that
radius.  This has the advantage of suppressing contamination from the wings of bright
stars and unmasked faint stars, while simultaneously offering a measure of the
photometric uncertainty in the total luminosity by comparison of the raw and
corrected aperture values.

Colors are determined through matched apertures, although this is very problematic at
small radii when comparing WISE filters to either ground-based optical or {\it
Spitzer} with their comparatively higher spatial resolution and superior PSF.  The rich
stellar fields at near-IR wavelengths make aperture to aperture comparison of WISE
images to other bandpasses difficult due contamination from the broad PSF wings of
foreground stars.  Extreme care is required to remove stellar contamination while not
reducing significant signal from the galaxy profile.  The use of matching metric
apertures at least offers some level of control of contaminating stars at large
radii.  For small apertures, the poor PSF in the WISE images makes aperture to
aperture comparisons impossible due to scattered light.

\subsection{The WISE and {\it Spitzer} Filter Systems}

WISE (Wide-field Infrared Survey Explorer) is a NASA 40-cm IR space telescope
launched into Sun-synchronous polar orbit in 2009.  Its primary mission was to survey
the sky at the mid-infrared bands of 3.4, 4.6, 12 and 22$\mu$m (Wright \etal 2010).
Our study is focused on the determination of the total baryonic mass of a galaxy and,
thus, WISE provides the IR photometry needed to equate luminosity to stellar mass
through the use of appropriate mass-to-light ratio ($\Upsilon_*$, see Cluver \etal
2014 and Kettlety \etal 2018).  The filter of
choice is 3.4 $\mu$m (W1) which uses a HAWAII 1-RG 1024$\times$1024 HgCdTe array with
a pixel scale of 2.75 arcsecs pixel$^{-1}$ producing a
47$^{\prime}$${\times}$47$^{\prime}$ field of view.  WISE images were extracted from
NASA/IPAC infrared science archive (IRSA) using the ALLWISE release.  Subimages
corresponding to 5 times the optical Holmberg radius were always adequate to insure
sufficient surrounding blank sky for background determination.  Rebinning by ALLWISE
resulting in a 1.375 arcsec pixel$^{-1}$ plate scale, with a slight improvement on
the PSF.

In comparison, the {\it Spitzer} Space Telescope was a 85-cm Ritchey-Chretien
launched into an Earth-trailing orbit in 2003 (Werner \etal 2004).  The primary
imaging instrument is the Infrared Array Camera (IRAC) which used an indium antimonide
256$\times$256 detector.  IRAC images were also obtained from IRSA using
the Enhanced Image archive.  Subimages varied on the field of view, but all had a
plate scale of 0.6 arcsecs pixel$^{-1}$ through channel 1 of the IRAC camera
(hereafter, IRAC 3.6).

\begin{figure}
\centering
\includegraphics[scale=0.85,angle=0]{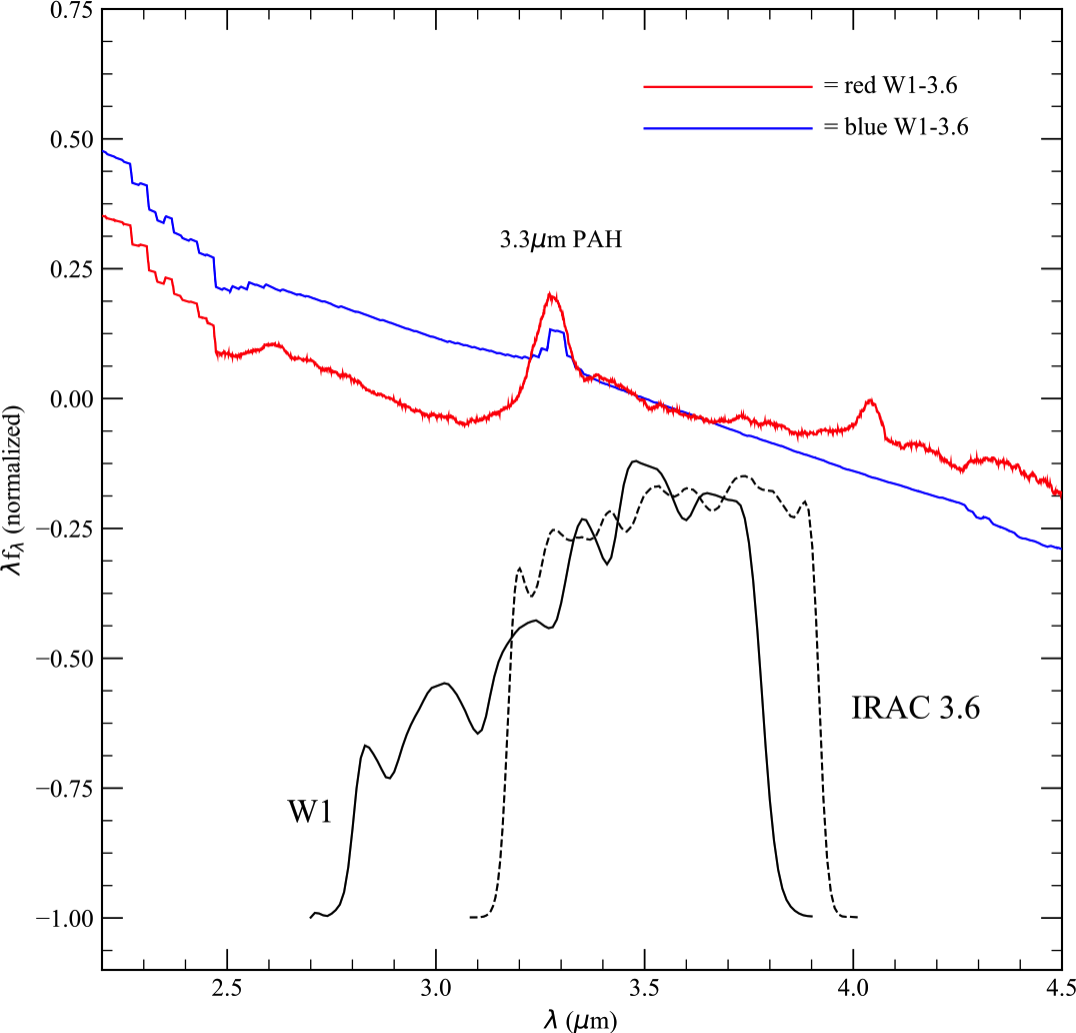}
\caption{\small The WISE W1 and {\it Spitzer} IRAC 3.6 filter responses in comparison
to two different mid-IR SED's from Brown \etal (2014).  Note that W1 and IRAC 3.6 have
nearly identical wavelength coverage where 73\% of the W1 flux matches 87\% of IRAC
3.6.  W1 samples its remaining flux to the blue of IRAC 3.6, and IRAC 3.6 samples its
remaining flux redward of 3.8$\mu$m.  Passive galaxies have steep SED's (shown as
blue) resulting in blue W1-3.6 colors.  Star-forming galaxies have flat SED's
resulting in red W1-3.6 colors (opposite to the traditional expectation for galaxy
colors).  Both filters encompass the 3.3$\mu$m PAH feature associated with AGB
stars.
}
\label{sed}
\end{figure}

A comparison of the W1 and IRAC 3.6 response curves are shown in Fig. \ref{sed}.
While the midpoints for W1 and IRAC 3.6 are listed as 3.37 and 3.55$\mu$m, the
filters are broad (from 2.8 to 3.9$\mu$m) with the W1 filter having a steeper slope
to the blue side.  The overlap region covers 73\% of the W1 response and 87\% of the
IRAC 3.6 flux.  Notably, both filters cover the PAH feature at 3.3$\mu$m, an emission
line associated with young AGB stars and loosely correlated with ongoing star
formation (see Meidt \etal 2012, Brown \etal 2014 and Querejeta \etal 2015).  Due to
the high overlap, the W1-3.6 color is expected to be near zero in the AB system.
However, there is a slight, but important color term that reflects underlying
astrophysics.  

Shown in Fig. \ref{sed} are two averaged spectral energy distributions (SED's) from
the Brown \etal (2014) sample.  Our choice of the Brown \etal dataset is one of
convenience as we have used this dataset in our earlier {\it Spitzer} studies.  While
more recent SED studies are available (see Clark \etal 2018), the Brown \etal SED's
are accurate over the wavelengths of interest herein.  An average of a dozen galaxies
with colors bluer than the mean W1-3.6 (see below) are shown in blue. For comparison,
an average with redder than mean colors are shown in red.  As discussed in the next
section, the bluer W1-3.6 galaxies are typically early-type in morphology, redder are
late-type.  The steeper SED reflects a stronger bulge component, which dominated the
IRS spectra.  Galaxies with strong star formation will have a flatter SED with a
stronger PAH 3.3$\mu$m feature.  The difference in colors can be seen as the effect
of a slightly longer red side for IRAC 3.6 (13\% outside the W1 band) and the wider
blue wing to W1 (27\% shortward of the IRAC 3.6 band).  Notice this results in a
reverse expectation for typical astronomical colors in that star-forming galaxies
will have redder W1-3.6 colors compared to quiescent galaxies with older stellar
populations.

It is also important to notice that a hot dust component due to increased star
formation only begins to be significant beyond 4 to 5$\mu$m.  While the 3.3$\mu$m PAH
feature increases with star formation, this is balanced in the W1 filter by a
relative decrease on the blue side of 3.3$\mu$m.  Thus, starlight dominates the flux
through both W1 and IRAC 3.6 and stellar mass estimates using these filters will only
have slight color component in the direction of higher $\Upsilon_*$ values for older
stellar populations (i.e. early galaxy types, see Schombert \etal 2022).  In
addition, the increased blue sensitivity to W1 compensates for stronger PAH emission
to produce a W1 luminosity conversion to stellar mass with much less distortion
due to star formation and AGB effects (Schombert \& McGaugh 2014).  However, a
detailed incorporation of an enhanced AGB contribution, as a component of a galaxy's
SED at 3.6$mu$m, is missing from most stellar population models in the literature
(see Conroy \& Gunn 2010) and we continue to use the empirical correction as a
function of metallicity from $K$ to 3.6 as given by Schombert \etal (2019).

\subsection{Comparison to WISE and {\it Spitzer} Photometry in the Literature}

\begin{figure}
\centering
\includegraphics[scale=0.85,angle=0]{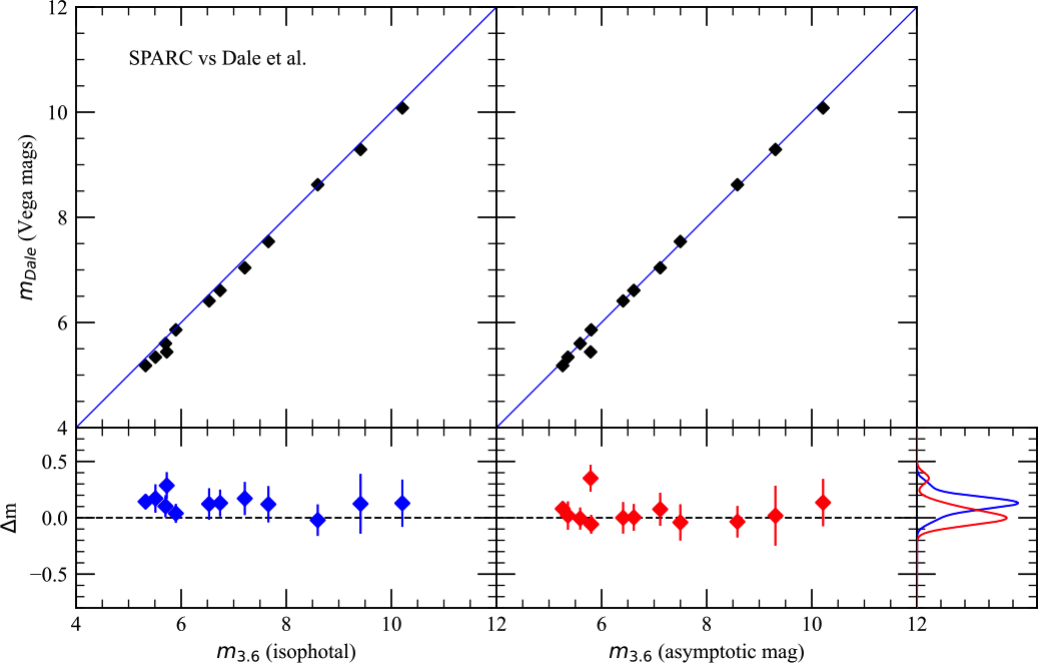}
\caption{\small A comparison between the Dale \etal (2007) IRAC 3.6 photometry
of the SINGS sample with the SPARC pipeline.  There were 12 galaxies in common, the
Dale \etal 3.6$\mu$m magnitudes are shown in comparison with SPARC isophotal
magnitudes (measured inside an elliptical aperture defined by the 23 3.6 mag
arcsecs$^{-2}$ isophote) and the asymptotic magnitude determined from curves of
growth.  The blue line represents unity between the magnitude systems, the
correspondence is excellent.  The residuals are shown in the bottom panels with a
normalized histogram for each panel to the far right.
}
\label{dale}
\end{figure}

One of the earliest near-IR space galaxy photometry projects was the {\it Spitzer}
Nearby Galaxies Survey (SINGS, Kennicutt \etal 2003).  The 3.6$\mu$m photometry was
presented in Dale \etal (2007) and used an optically determined elliptical aperture.
An additional extended source aperture correction was applied to account for
scattered light, typically in the 8 to 9\% range.  There were 12 galaxies in common
from the SINGS sample to the SPARC sample, and the resulting $m_{3.6}$ values (in the
Vega system) are shown in Fig. \ref{dale}.  Two comparisons with the SPARC
photometry are shown, the left panel displays the isophotal magnitude
from our photometry pipeline that measures the luminosity inside an elliptical
aperture defined by the 23 mag arcsecs$^{-2}$ isophote (this value is selected as it
roughly corresponds to the Holmberg radius of 26.5 $B$ mag arcsecs$^{-2}$ for the
color of late-type galaxies).  The right panel displays the 3.6 $\mu$m asymptotic
magnitude determined from curves of growth described above.  

In general, the correspondence is excellent. The differences between the Dale \etal
luminosities and SPARC are less for the asymptotic magnitudes, but the shift is minor
(less than 0.05 mags).  The aperture correction procedure for Dale \etal seems to
recover all the missing outer flux and is in good agreement with SPARC's
determination of a total luminosity by curves of growth.  The typical internal error
is quoted as 0.01 by Dale \etal (2007) but the dispersion with respect to the
SPARC asymptotic magnitudes is on the order of 0.05 mags.

The most extensive {\it Spitzer} galaxy photometry survey is the S$^4$G project
(Sheth \etal 2010) which observed 2,331 galaxies.  The techniques used by the S$^4$G
pipeline are very similar to the ARCHANGEL system used by the SPARC pipeline.
Background determination used sky boxes and elliptical apertures are defined by the
outer isophotes of the target galaxy.  Total luminosities are assigned by fitting the
accumulated magnitude as a function of the magnitude gradient to obtain an asymptotic
magnitude.  The only minor difference between SPARC and S$^4$G pipelines is the lack
of replacement of masked pixels with a nearby mean intensity, which typically results
in a systematic shift of 5\% for {\it Spitzer} elliptical samples (see Schombert \&
McGaugh 2014).

\begin{figure}
\centering
\includegraphics[scale=0.85,angle=0]{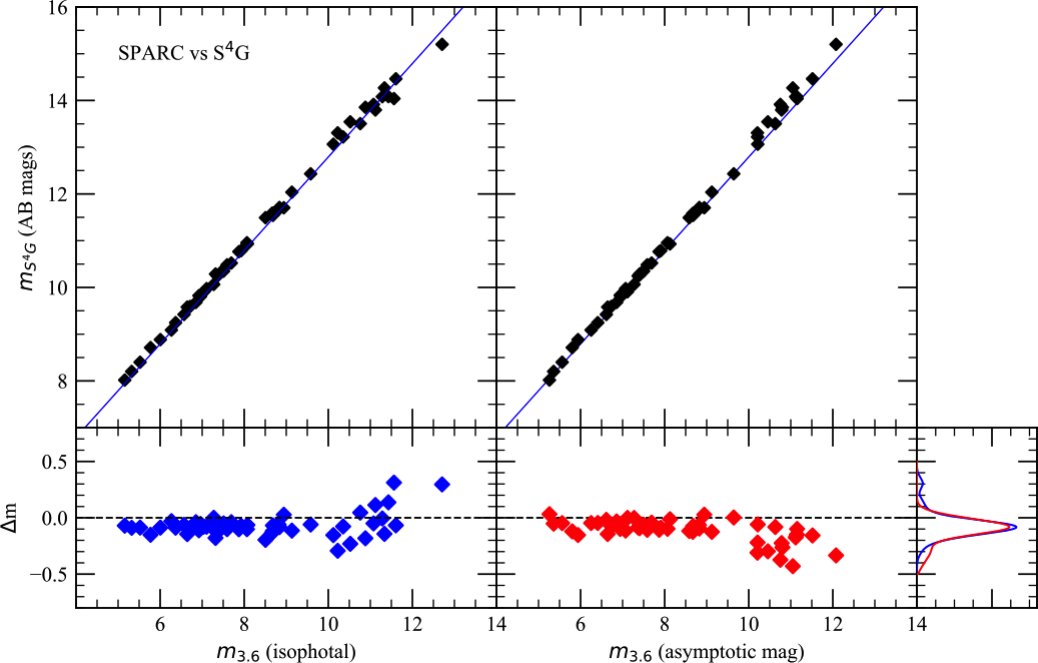}
\caption{\small A comparison between the S$^4$G IRAC 3.6 photometry (Sheth \etal
2010) with the SPARC pipeline.  There were 51 galaxies in common, the correspondence
is good with a 7 to 10\% shift between SPARC and S$^4$G fluxes with either isophotal
apertures or asymptotic magnitudes.  The residuals are shown in the bottom panels with a
normalized histogram for each panel to the far right.
We attribute the difference to varying
photometric techniques (see text). 
}
\label{s4g}
\end{figure}

There were 51 galaxies in common with SPARC and S$^4$G samples, their comparison is
shown in Fig. \ref{s4g}.  While the correspondence is one-to-one, the S$^4$G
luminosities were 10\% fainter than our asymptotic magnitudes and 7\% fainter than
our isophotal magnitudes.  This does not reflect uncertainties in the photometry but
rather the different techniques used to determine a total luminosity.  Most of the
difference is found at the faint magnitudes where our procedure of replacing masked
pixels with mean galaxy surface brightness produces a notable increase in total
galaxy luminosity compared to the brighter galaxies.  This also serves as a
cautionary tale considering the error budget for galaxy photometry is often quoted at
the 1\% level, but clearly the technique and isophote chosen can lead to 10\%
differences in various determinations of a total stellar mass.  The dispersion around
the offset magnitude is on the order of 0.08 mags.

With respect to WISE W1 photometry, one of the earliest papers to use WISE photometry
for analyzing the Tully-Fisher relation is Neill \etal (2014).  Using WISE W1 and W2
photometry on 310 galaxies in 13 clusters, they explore the absolute magnitude to \hi
line width relationship. As with the S$^4$G project, their photometry pipeline is
similar to SPARC using drizzled WISE images extracted from IRSA.  Total luminosities
are defined from an asymptotic magnitude that is the integration of the galaxy radial
profile.

\begin{figure}
\centering
\includegraphics[scale=0.85,angle=0]{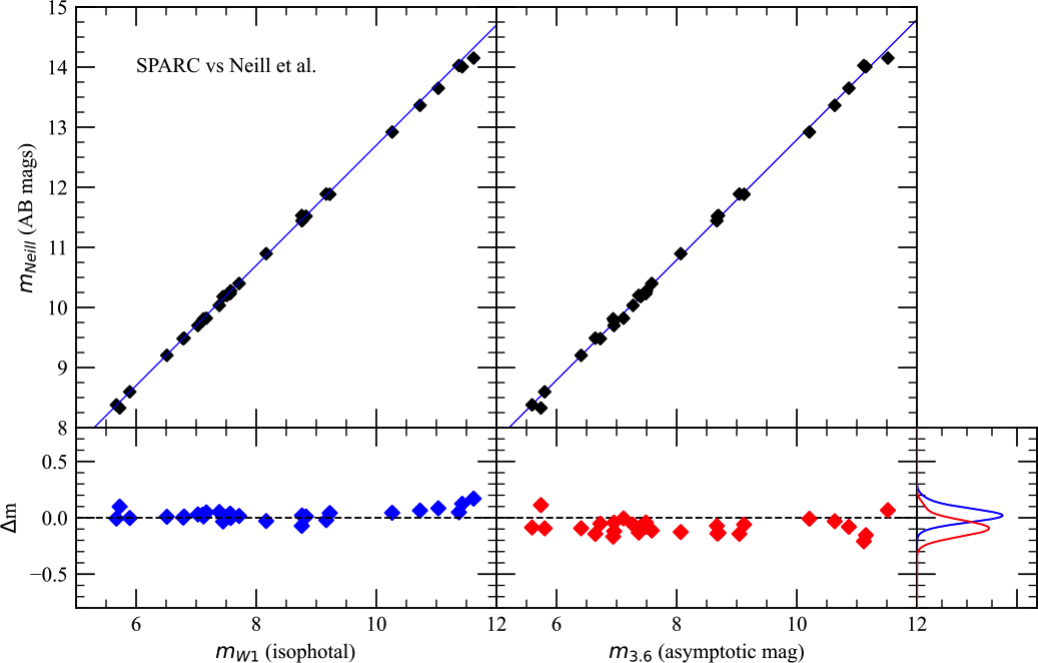}
\caption{\small A comparison between the Neill \etal WISE photometry and SPARC WISE.
Both SPARC and Neill \etal use similar photometric pipelines, particularly with
respect to the use of drizzled WISE images.  The residuals are shown in the bottom panels with a
normalized histogram for each panel to the far right. The correspondence is excellent with a
slight deviate at low luminosities.  The comparison with asymptotic 3.6 magnitudes
recovers this missing flux.
}
\label{neill}
\end{figure}

The SPARC and Neill \etal samples have 28 galaxies in common, shown in Fig.
\ref{neill}.  The agreement is excellent with a slight tendency for the SPARC
isophotal W1 values to underestimate the Neill \etal asymptotic magnitudes at faint
luminosities (roughly 5\% below $m_{W1}$ = 10).  The comparison to {\it Spitzer}
3.6$\mu$m asymptotic is also strong with a mean offset of 0.1 mags in concurrence
with the average $W1-3.6$ color term.  Again, the mean internal photometry error at
W1 is similar to 3.6 at the 0.01 mag level, although the dispersion is on order of
0.1 mags.

\begin{figure}
\centering
\includegraphics[scale=0.85,angle=0]{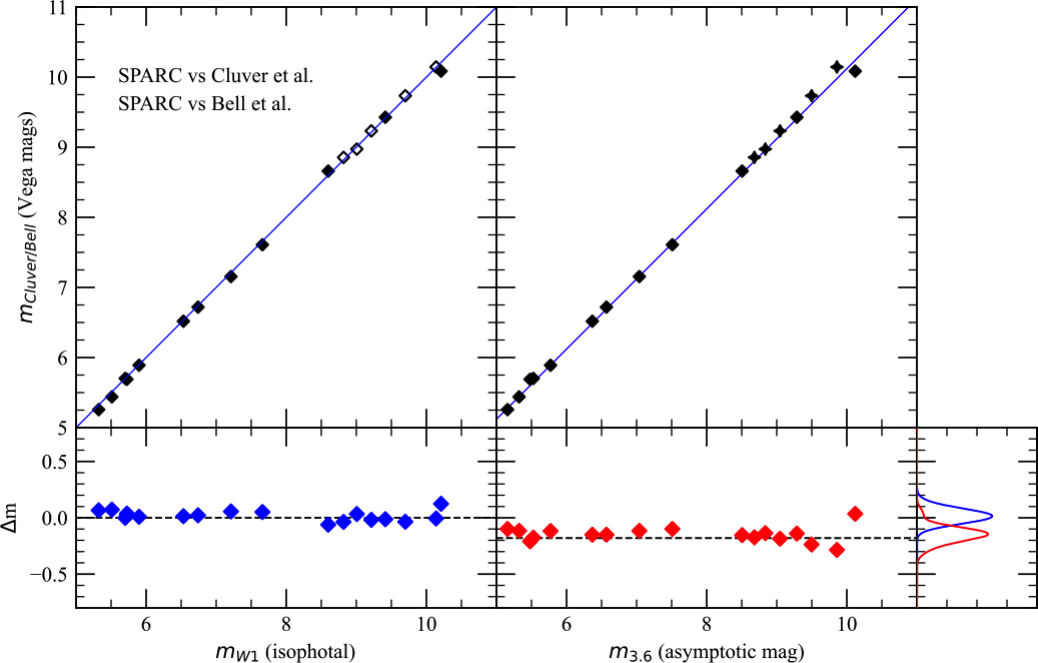}
\caption{\small A comparison with the W1 and IRAC 3.6 photometry of Cluver \etal
(2014) and Bell \etal (2022).  The residuals are shown in the bottom panels with a
normalized histogram for each panel to the far right. Both these studies use the procedures developed in
Jarrett \etal (2013) and there is an excellent correspondence over five orders of
magnitude.
}
\label{cluver}
\end{figure}

Two other recent surveys of the WISE images related to the Tully-Fisher relation are
Cluver \etal (2014) and Bell \etal (2022).  Both studies follow the procedure
pioneered in Jarrett \etal (2013) which is nearly identical to the SPARC photometry
pipeline in using elliptical apertures with curves of growth combined with replacing
masked pixels with local intensity values.  These studies have 17 galaxies in common
with the SPARC WISE sample and are shown in Fig. \ref{cluver}.  As before, the
agreement is excellent with a dispersion of only 0.08 mags.  While this is higher
than the quoted errors in all the photometry pipelines, it more accurately reflects the
limit to repeatability in galaxy photometry.  The {\it Spitzer} asymptotic magnitudes
display the same correspondence with a color offset of 0.2 mags.

\begin{figure}
\centering
\includegraphics[scale=0.85,angle=0]{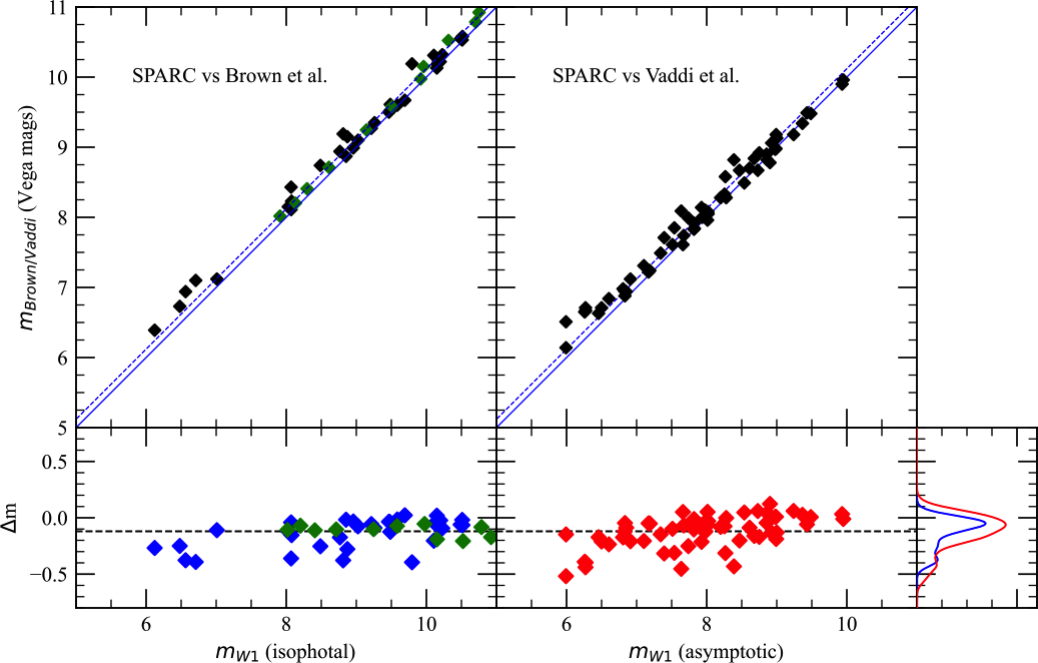}
\caption{\small A comparison with the SED study of Brown \etal (2014), Clark \etal
(2018) and WISE
photometry from Vaddi \etal (2016).  The residuals are shown in the bottom panels with a
normalized histogram for each panel to the far right. The correspondence is good, but the dispersion
is higher than previous comparisons.  For the Brown \etal (black symbols) and Clark
\etal (green symbols) studies, this reflects
varying aperture sizes that appear to reproduce a Kron magnitude (90\% the total
luminosity shown by the dashed line).  The Vaddi \etal also has a larger dispersion,
but with the largest overlap to the SPARC sample in terms of galaxy numbers.  The dashed line is the 90\%
flux comparison.  The dispersion from this offset is 0.06 mags, which probably
reflects the true uncertainty in galaxy photometry regardless of error bars quotes by
each study.
}
\label{vaddi}
\end{figure}

Lastly, we compare our pipeline results with two spectrophotometric surveys, Brown
\etal (2014) and Vaddi \etal (2016).  The comparison with Brown \etal is particular
salient as that study was a comprehensive analysis of the SED's of 129 galaxies from
the UV to the mid-IR for varying morphological types and a range of current SFR's.
There were 34 galaxies in common with an ongoing extension of the SPARC sample (SPARC
1k) and comparison is shown in the left panel of Fig. \ref{vaddi}.  The focus of
Brown \etal was to match standard filter flux to the deep optical and IR spectra for
the same galaxies.  This resulted in different apertures for their W1 and 3.6$\mu$m
luminosities, which would seem to explain the large scatter in Fig. \ref{vaddi}.
We note that the offset goes to zero if we assume the Brown \etal fluxes are based on
a Kron magnitude definition (i.e., 90\% the total flux, shown by the dotted line in
Fig. \ref{vaddi}), but the dispersion is still larger than accountable by pure
photometric errors.  Also shown are the aperture magnitudes from the DustPedia survey
(Clark \etal 2018), a similar SED study from the UV to far-IR with 11 galaxies in
common with our SPARC sample.

The data from Vaddi \etal had 61 galaxies in common with the new SPARC 1k sample.
Vaddi \etal use an isophotal defined aperture where the aperture size is given as 
one standard deviation above mean sky value.  Foreground stars are masked, but
missing flux is not replaced.  Again, as seen in the right panel of Fig.
\ref{vaddi}, the dispersion is higher than other samples, and the SPARC sample are
typically 5 to 10\% brighter than the Vaddi \etal fluxes. Again, we interpret this
difference as due to technique differences rather than errors in the photometry.

\subsection{W1 to IRAC 3.6 colors}

The WISE to {\it Spitzer} color sequence is poorly explored in the literature.  This
is primarily due to the fact that since W1 and IRAC 3.6 are so similar in their
wavelength response, the two fluxes are nearly identical and there was limited
information in their ratios with respect to star formation history.   However, there
is a clear distinction between blue and red W1-3.6 values with respect to the Brown
\etal SED's.  The slight blue and red wings to the W1 and IRAC 3.6 filters maps well
into the two distinct SED shapes for galaxies; 1) power-law shaped SED and 2) ones
with flat flux around the 3.3$\mu$m PAH feature.  In particular, we found in \S2 that
steep SED's are associated with blue W1-3.6 colors and flat SED's are aligned with
red W1-3.6 colors.  However, the difference between the blue and red SED's in Fig.
\ref{sed} is only 10\% of the total W1 flux, which corresponds to a small shift of
0.2 mags in the AB system.

\begin{figure}
\centering
\includegraphics[scale=0.85,angle=0]{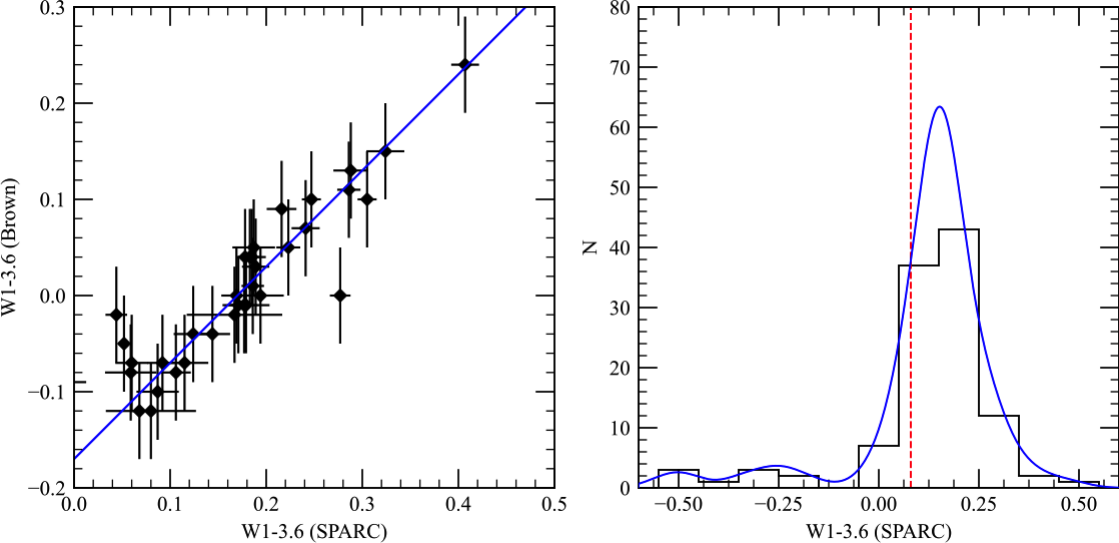}
\caption{\small The left panel displays a comparison of aperture colors between SPARC
and Brown \etal.  The relationship is one-to-one in relative values, and the
dispersion is within the observational errors (a mean error bar is shown).  However,
there is a 0.17 mag shift from the Brown \etal W1-3.6 colors compared to SPARC.  We
attribute this to differences in the zeropoint of the AB system used by Brown \etal
and varying aperture sizes between their WISE and {\it Spitzer} measurements.  The
right panel displays a histogram of SPARC W1-3.6 colors.  We note that a majority are
redward of the mean elliptical color (red dashed line, Schombert 2016), in agreement with the expectation
that star-forming disks will have redder colors than passive, old stellar population
galaxies.
}
\label{color}
\end{figure}

There are few studies in the literature with matching WISE and {\it Spitzer}
apertures to compare colors.  The Brown \etal sample provided adjusted W1 and IRAC
3.6 fluxes for comparison to SPARC W1-3.6 colors.  This comparison is shown in Fig.
\ref{color}.  The correspondence is good except for a 0.17 mag shift due to a
calibration shift from AB to Vega magnitudes plus a mismatch between the WISE and
IRAC 3.6 apertures in the Brown \etal sample.  However, the dispersion from a line of
unity matches the typical photometric errors in each filter, giving some confidence
that both techniques are measuring the same flux difference, i.e., the external color
errors are in the range of 0.04 to 0.06 mags.

A subsample of the SPARC sample was selected for color analysis where each galaxy was
without contaminating bright stars near the isophotal aperture.  This resulted in 111
galaxies with good W1-3.6 values shown as a histogram in Fig. \ref{color}.  The mean
W1-3.6 color of a $L^*$ elliptical is shown as the red line (Schombert 2018).  The
SPARC sample primarily falls in the red side of W1-3.6 colors, which we have seen in
the previous section associates with SED's dominated by recent star formation.  This
agrees with the fact that SPARC focuses on a sample of primarily spirals and dwarfs
irregulars with varying amounts of ongoing star formation.  The early-type galaxies,
with prominent bulges, occupy the blue side of the histogram.  Galaxies with Seyfert
or LINER signatures have the bluest W1-3.6 colors (see the next section).

\section{The WISE SPARC Sample}

A subset of 111 galaxies were selected from the 175 galaxies in the original SPARC
sample.  The subset was chosen for good image quality in both WISE and {\it Spitzer}
frames, meaning a well-defined galaxy profile with no nearby bright stars or bright
galaxies. In order to ensure that resolution differences between WISE and {\it
Spitzer} did not bias the aperture magnitudes, we placed a 30 arcsec radius limit to
the metric apertures. 

The SDSS, WISE and {\it Spitzer} magnitudes are listed in Table 1, where the aperture
used was the ellipse from {\it Spitzer} surface photometry interpolated to the 23
3.6$\mu$m mag arcsecs$^{-2}$ isophote.  In addition to WISE and {\it Spitzer}
photometry, the SDSS archive was searched for SDSS $g$ and $r$ images for the same
galaxies.  Of the WISE sample, 79 were found in the SDSS archive.  The same metric
aperture was applied to the SDSS images.

Errors were assigned from two characteristics of the images.  The first is standard
Poisson noise from the galaxy flux itself.  All the galaxies were significantly
larger than either the {\it Spitzer} or WISE PSF's, so the areal flux was converted
back into photon counts for a $\sqrt{N}$ determination.  The Poisson error was always
a factor of ten smaller than the error due to the background noise.  The error in the
sky value was assigned through the use of sky boxes. Typically over ten boxes of 20
by 20 pixels were selected from regions at the edge of the frames devoid of stars or
faint galaxies.  The pixels in each box were averaged with a jack-knife procedure and
the dispersion of the mean of those boxes is assigned as the error on the sky value.
The area of the aperture is multiplied by sky error and added in quadrature with the
Poisson error for the final uncertainty in magnitude listed in Table 1.

\begin{figure}
\centering
\includegraphics[scale=0.85,angle=0]{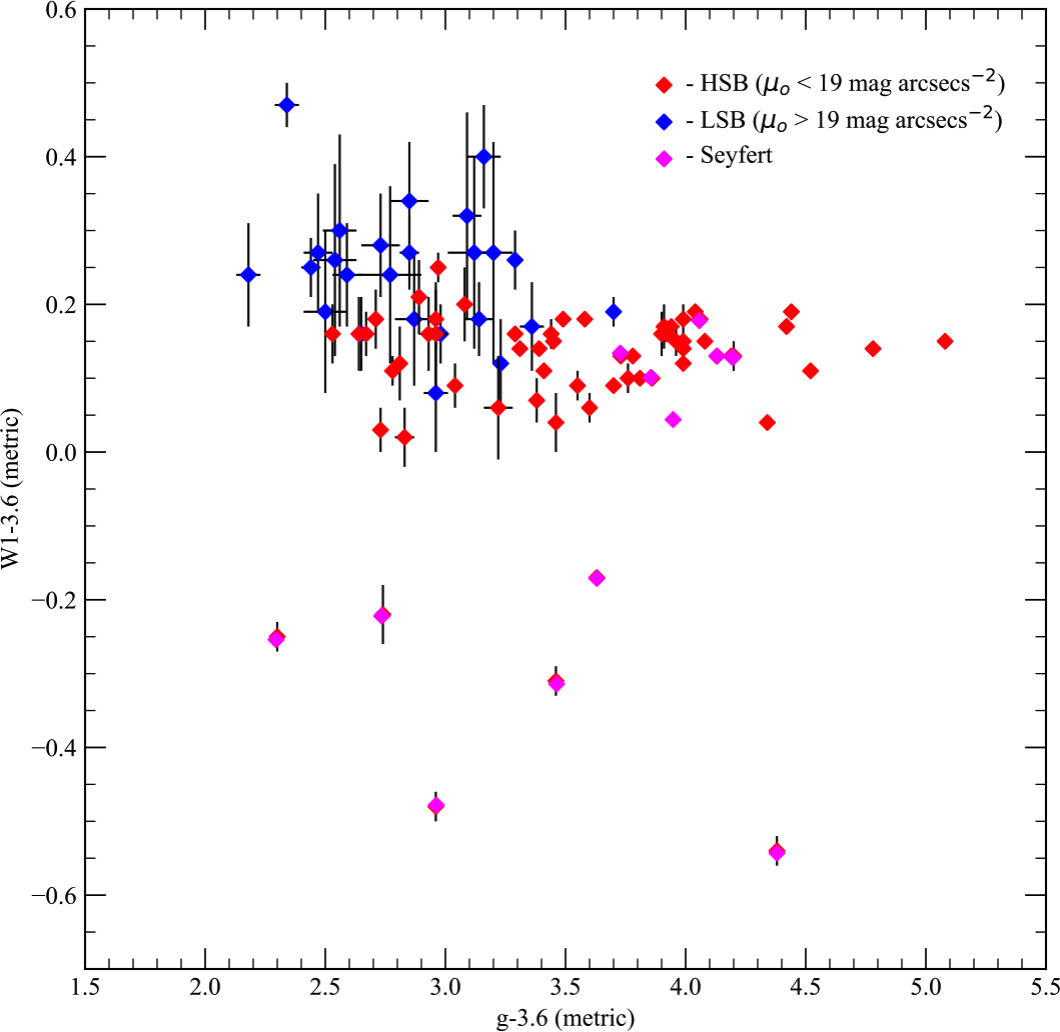}
\caption{\small The optical $g$-3.6 color compared to the IR W1-3.6 color for 87
galaxies in the SPARC sample with both WISE and SDSS imaging.  While there is a wide
dispersion in optical color, the range in W1-3.6 color is quite small (when Seyferts
are excluded). The distinction between LSB and HSB galaxies in W1-3.6 probably
reflects a metallicity effect (see Fig. \ref{cmr}).
}
\label{two_color}
\end{figure}

The resulting $g$-3.6 versus W1-3.6 two color diagram from Table 1 is shown in Fig.
\ref{two_color}.  The $g$-3.6 color is similar to $V$-3.6 used in our previous star
formation history studies (Schombert \etal 2019) and covers the full range in color
found for spirals and irregulars noted in larger galaxy catalogs.  In particular,
$V$-3.6 is the benchmark color for assigning a mass-to-light ratio ($\Upsilon_*$)
value from stellar population models (see Fig. 2, Schombert \etal 2022). 

The general characteristics of the two color diagram followed those found for the
$K$-3.6 diagram (Schombert \& McGaugh 2014).  There is a slight redward slope, which
we now understand as due to flatter SED's at 3.6$\mu$m for star-forming, optically
blue galaxies.  There are several galaxies with W1-3.6 values below zero that signal
steep SED's associated with quiescent stellar populations.  However, we note that all
the very blue W1-3.6 galaxies (W1-3.6 less than 0) also show Seyfert or LINER
signatures in their optical spectra.  Seyferts lack the 3.3$\mu$m PAH feature in the
Brown \etal SED's combined with much steeper blueside fluxes in the W1 filter.  This
non-thermal component would explain the extreme W1-3.6 colors.

We also note the unusual color signature displayed by low surface brightness (LSB)
galaxies.  If the SPARC sample is divided in high and low central surface brightness
(defined in Schombert \& McGaugh 2014), then a clear color distinction is displayed
such that LSB galaxies have redder W1-3.6 colors on average (also bluer $g$-3.6
colors, which was well known from early LSB studies, Pildis \etal 1997).  Unusual
colors for LSB disks is usually interpreted as a signature of low metallicities
(Schombert \& McGaugh 2021) and lower metallicity does drive up the redside of the
IRAC 3.6 filter, based on $K$-3.6 colors and stellar population models (see Schombert
\etal 2022).

The distinction between LSB and HSB galaxies in W1-3.6 color can be seen in the
stellar population models of Schombert \etal (2019), where the steady decline in
$\Upsilon_*$ levels off at $g$-3.6 colors less than 3.  The increased flux in the
IRAC 3.6 filter balances what would normally be the mass-to-light ratios at those
wavelengths.  Interestingly, the sample of HSB galaxies, with Seyfert objects
removed, have a nearly constant W1-3.6 color of 0.14$\pm$0.05.

\begin{figure}
\centering
\includegraphics[scale=0.85,angle=0]{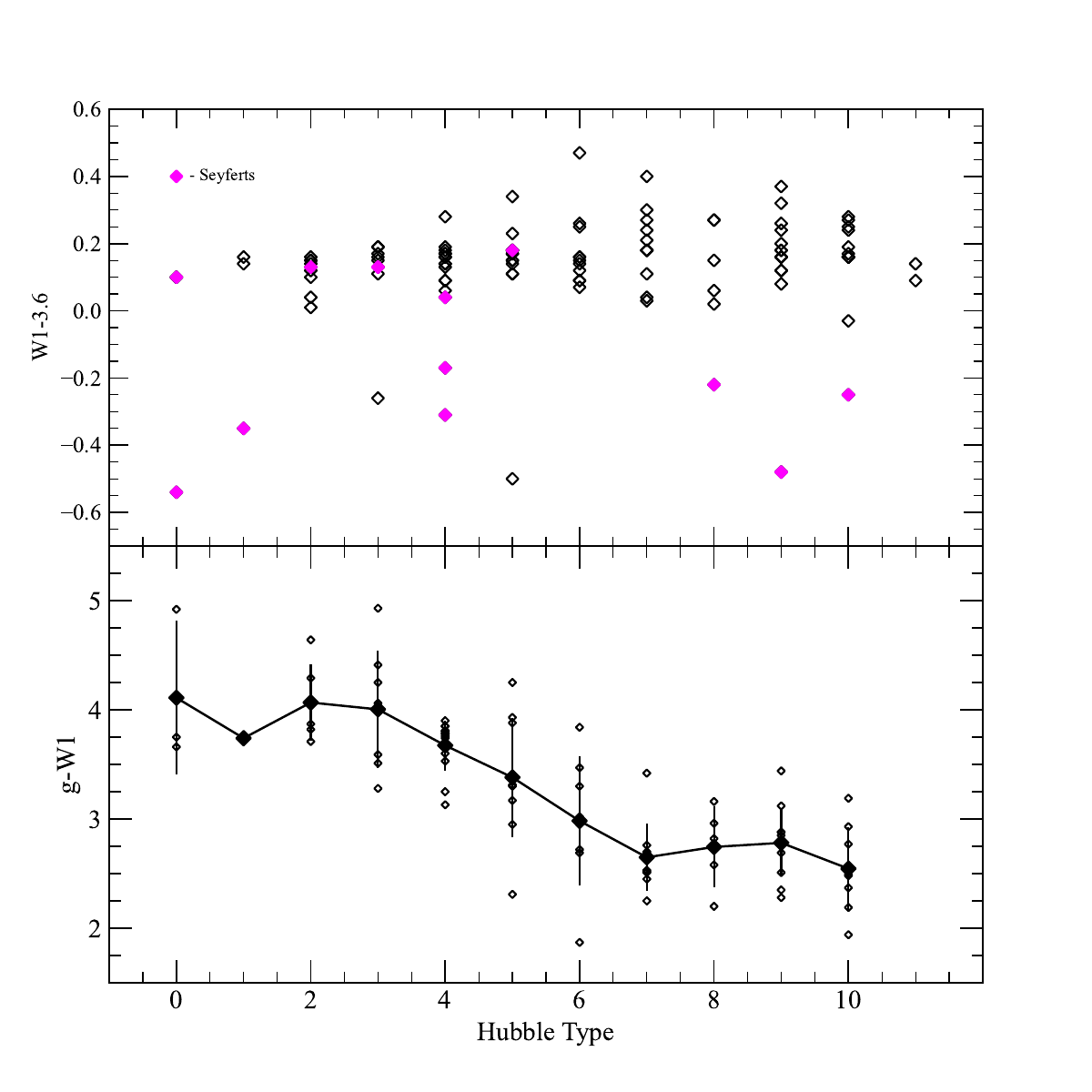}
\caption{\small The range in optical and IR color with respect to Hubble type for the
SPARC sample.  The trend in $g$-W1 matches the known trends outlined in Schombert \&
McGaugh (2021).  The slight trend for redder W1-3.6 color with later Hubble type is not
statistically significant and demonstrates why IR fluxes are significantly better for
stellar mass determinations than optical flux due to their independence from star
formation history as related to galaxy type.
}
\label{hubble}
\end{figure}

The meaning of the W1 and IRAC 3.6 colors is clearer when we compare W1-3.6 and the
SDSS $g$-3.6 colors as a function of Hubble type (shown in Fig. \ref{hubble}).  The
correlation of bluer optical colors with later Hubble types is clear, the colors
closely correspond to the $V$-3.6 colors from Schombert \etal (2019).  In fact, as
argued in Schombert \etal (2019), the trend in Hubble type with color also tracks the
deduced $\Upsilon_*$ model value such that either color or morphology serves equally
well to assign an accurate $\Upsilon_*$ to a particular galaxy.  Thus, the clear
trend with $g$-W1 is reassuring as this color is used to define the
color-$\Upsilon_*$ relationship from stellar population models.

The trend of W1-3.6 with Hubble type is distinctly different.  There is a slight
tendency to find redder W1-3.6 colors with later Hubble types, but a least-squares fit
is indistinguishable from a zero slope line.  The dispersion in color is similar for
each category from $g$-W1 to W1-3.6.  Again, we note extremely blue W1-3.6
galaxies have Seyfert signatures.  We expect the stellar mass values assigned from
either W1 or IRAC 3.6 fluxes will produce similar values, unless the galaxy in
question displays AGN activity.

\begin{figure}
\centering
\includegraphics[scale=0.85,angle=0]{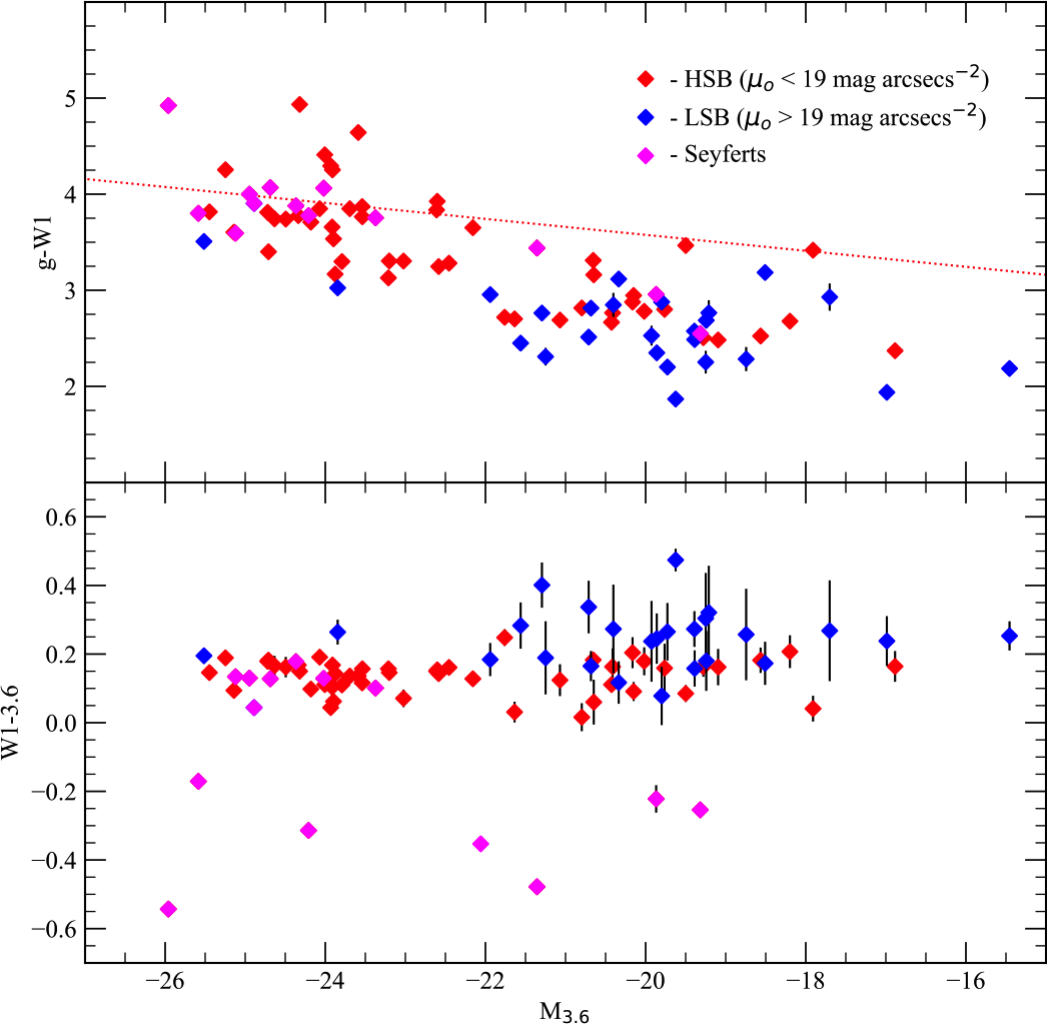}
\caption{\small The near-IR and optical color-magnitude relation for the SPARC
sample.  The upper panel displays the SDSS $g$ versus WISE W1 color where HSB and LSB
are marked as red and blue respectively.  Seyferts or LINER galaxies are colored
magenta.  The CMR for ellipticals are indicated by the red dotted line (Schombert
2018).  The difference between the passive ellipticals and spirals/dwarfs of the
SPARC sample is
evident in the steeper, bluer colors due to recent star formation. The dispersion in
the elliptical CMR is very small at 0.1 mags whereas the large scatter 
in the SPARC galaxies represents a wide range in star formation histories as
well as slower chemical enrichment.  The lower panel displays the CMR for the near-IR color
W1-3.6 where star formation effects are negligible.  Aside from the Seyfert outliers, the range in W1-3.6 color is small and
constant over the luminosity range of the SPARC sample.
}
\label{cmr}
\end{figure}

Lastly, the most common color correlation with respect to galaxies is the
color-magnitude relation (CMR), usually interpreted to be an age/metallicity
correlation between galaxy mass and color (see Sextl \etal 2023 for a recent review).
Fig. \ref{cmr} displays the optical and IR CMR with respect to absolute 3.6$\mu$m
luminosity.  The correlation with optical $g$-W1 color is clear and with a scatter
typical for late-type galaxies (reflecting varying star formation history paths,
Tojeiro \etal 2013).  The dotted line displays the CMR for pure ellipticals
(Schombert 2018) that track the mass-metallicity relationship for single burst
stellar populations.

As shown in Tojeiro \etal (2013), passive galaxies display IR colors proportional to
their stellar mass in agreement with a standard chemical enrichment scenario where a
greater gravitational potential produces rapid grow in metallicity and redder stellar
populations.  They also found that early-type spirals, where the galaxy color is
dominated by metal-rich bulges, also overlap the elliptical colors.  Galaxies with
more star formation display bluer $g$-W1 colors and form the so-called 'blue cloud'
below the elliptical sequence.  Galaxies below $-$22 divide into two groups where
lower mean metallicity pushes LSB galaxies to bluer colors than expected from their
star formation histories.

The IR W1-3.6 color displays a slight inverse correlation, mostly driven by LSB
galaxy colors, that tracks the expectation of metallicity models for low mass dwarf
galaxies.  Again, Seyfert galaxies with their steep SED's are indicated and form 80\%
of the outliers.  The consistence in the near-IR colors again demonstrates an
important aspect with respect to stellar mass determination in that neither strong
star formation nor a wide range in metallicity alter the W1 versus IRAC 3.6 fluxes by
a significant amount.  Aside from AGN activity, the near-IR fluxes are remarkably
stable over a range of galaxy colors and morphological types.

\section{Summary}

The primary goal of this study was to link the {\it Spitzer} and WISE photometry
results for the SPARC sample with special attention to the role of near-IR colors for
stellar population models that are used to estimate the stellar mass component of a
galaxy.  We summarize are main results as the following:

\begin{itemize}

\item[(1)] The photometry pipeline used for this, and future, SPARC projects focuses
on the use of asymptotic magnitudes for total luminosities and metric apertures for
colors.  Notable differences between {\it Spitzer} and WISE photometric values were
found to be due to the much poorer PSF for the ALLWISE archive compared to {\it
Spitzer}.

\item[(2)] A comparison of several WISE W1 and IRAC 3.6 datasets confirms that
differing photometry techniques results in variations of up to 10\% for galaxies in
common.  When comparing with studies that use identical photometry techniques the
differences are at the 0.01 mag level, which probably represents the true error in
galaxy total luminosities, rather than the formal errors output from the various
pipelines.

\item[(3)] The W1-3.6 colors display excellent one-to-one correspondence with SED
studies (e.g., Brown \etal 2014).  Positive W1-3.6 colors are associated with galaxies
with significant star formation.  Strongly negative W1-3.6 colors are found for
Seyfert and LINER galaxies.  For non-AGN system, the range in W1-3.6 color is fairly
limited and only slighter redder than the mean for elliptical galaxies.

\item[(4)] The optical to near-IR two-color diagram indicates a notable difference
between LSB and HSB galaxies, probably related to mean metallicity differences (i.e.,
bluer RGB populations) rather than star formation as their $g$-3.6 colors are
similar.  We found that correcting for central surface brightness will improve the
stellar mass estimates by 5\%, depending on the suite of stellar population models
used.

\item[(5)] The W1-3.6 color is independent of morphology type, other than the
tendency for LSB galaxies to be of late Hubble types, while the optical $g$-W1 color
is very sensitive to galaxy type.  The distinction will be important in linking WISE
to {\it Spitzer} stellar mass models in our second paper.

\item[(6)] The optical and near-IR color-magnitude diagrams display the same blue
versus red cloud features noted by earlier studies, particularly the SDSS colors of
Tojeiro \etal (2013).  The $g$-W1 CMR reproduces all the features of the SPARC
$g$-3.6 CMR and, in addition, singles out galaxies with strong AGN features.  The
near-IR W1-3.6 CMR display almost zero slope, raising our confidence in applying
mass-to-light ratios across a range of stellar masses.

\end{itemize}

\section*{Acknowledgements} Software for this project was developed under NASA's AIRS
and ADAP Programs.  This work is based in part on observations made with the Spitzer
Space Telescope, which is operated by the Jet Propulsion Laboratory, California
Institute of Technology under a contract with NASA.  Support for this work was
provided by NASA through an award issued by JPL/Caltech. Other aspects of this work
were supported in part by NASA ADAP grant NNX11AF89G and NSF grant AST 0908370. As
usual, this research has made use of the NASA/IPAC Extragalactic Database (NED) which
is operated by the Jet Propulsion Laboratory, California Institute of Technology,
under contract with the National Aeronautics and Space Administration.  FD wishes to
thank the University of Oregon's Presidential Undergraduate Research program which
supported the initial stages of this project as her honors thesis.

\begin{deluxetable}{lccccc}
\tablecolumns{6}
\small
\tablewidth{0pt}
\tablecaption{SPARC Photometry}
\tablehead{
\\
\colhead{Name} & \colhead{R} & \colhead{m$_g$} & \colhead{m$_r$} & \colhead{m$_{W1}$} & \colhead{m$_{3.6}$} \\
 & \colhead{(arcsecs)} & & & & \\
}
\startdata
F568-1         & 28.3 & 16.173$\pm$0.102 &  ...             & 13.864$\pm$0.153 & 13.675$\pm$0.148 \\
F568-3         & 42.5 & 15.826$\pm$0.107 & 15.411$\pm$0.092 & 13.376$\pm$0.072 & 13.093$\pm$0.114 \\
F568-V1        & 21.1 & 16.705$\pm$0.114 & 16.345$\pm$0.097 & 14.192$\pm$0.091 & 13.855$\pm$0.123 \\
F571-V1        & 15.6 & 17.680$\pm$0.160 & 17.288$\pm$0.149 & 15.152$\pm$0.130 & 14.915$\pm$0.196 \\
F574-1         & 30.2 & 16.777$\pm$0.096 & 16.333$\pm$0.073 & 14.014$\pm$0.066 & 13.613$\pm$0.114 \\
\enddata
\tablecomments{Only the first five galaxies are shown.  The rest are in the electronic version.}
\end{deluxetable}

\end{document}